# Transporting of a Cell-Sized Phospholipid Vesicle Across Water/Oil Interface

**Masahiko Hase, Ayako Yamada, Tsutomu Hamada and Kenichi Yoshikawa**[*]

*Department of Physics, Graduate School of Science, Kyoto University, Kyoto 606-8502, Japan*

[*]*to whom correspondence should be addressed. E-mail: yoshikaw@scphys.kyoto-u.ac.jp*

**Abstract**

When a cell-sized water droplet, with a diameter of several tens of μm, is placed in oil containing phospholipids, a stable cell-sized vesicle is spontaneously formed as a water-in-oil phospholipid emulsion (W/O CE) with a phospholipid monolayer. We transferred the lipid vesicle thus formed in the oil phase to the water phase across the water/oil interface by micromanipulation, which suggests that the vesicle is transformed from a phospholipid monolayer as W/O CE into a bilayer. The lipid vesicle can then be transported back into the oil phase. This novel experimental procedure may be a useful tool for creating a model cellular system, which, together with a microreactor, is applicable as a micrometer-scale biochemical reaction field.

1. Introduction

The cytoplasmic membrane is constituted with phospholipid, which is a surface active molecule.[1] In aqueous solution, phospholipids spontaneously form closed bilayer phospholipid vesicles, referred to as liposomes, which have their acyl chains face-to-face. Liposomes have been often used as a cell model and as a microreactor for micro- or submicroscale biochemical reactions.[2-4] Currently, the most popular method for preparing liposomes is sonication.[5] With this method, small liposomes on the order of several tens of nm are obtained, which indicates that the curvature of the membrane is much greater than that of natural biomembranes. Recently, cell-sized liposomes (CLs) on the order of several tens of μm have been actively studied. CLs are stable in response to thermal fluctuation due to their low curvature, and large enough for real-time manipulation and observation by optical microscopy.[6,7] Some studies have been performed on their use as a cell model, e.g. on the encapsulation and functional expression of biochemical substances such as gene and/or protein[8-12], and on their application in a microreactor.[13,14] Several methods for preparing CLs have been reported, including reversed-phase evaporation[15], solvent vaporization[16], preparation with detergent[17], fusion with $Ca^{2+}$ for negatively charged lipid[18], swelling with water of almost zero ionic or ionic strength[19-21], swelling of charged lipid with an aqueous solution,[22] and combinations of the above techniques. However, so far it has been difficult to prepare CLs that incorporate substances from the outer environment or CLs with an asymmetric distribution of phospholipids between their inner and outer membranes, like cytoplasmic membranes. In addition, it is difficult to fuse CLs to each other, which limits the applicability of CLs as microreactors.

On the other hand, a cell-sized water droplet in oil containing phospholipid forms a stable cell-sized vesicle with a phospholipid monolayer at the water-oil boundary as a water-in-oil phospholipid emulsion (W/O CE)[23], and has been expected to be a prospective tool for a microreactor, since it is easy to control the size, fusion, and encapsulation of substances.[24-26] Additionally, W/O CEs can be considered to be a cell model because the hydrophilic parts of phospholipids face the water phase at the water/oil boundary, as in a cytoplasmic membrane.

If the transition of a phospholipid vesicle between W/O CE and CL can be easily achieved, a lipid vesicle may serve as a better cell model and a high-performance microreactor. Very recently, preliminary studies suggested that CLs can be formed from W/O CE across a water/oil interface[11,27], however direct observation of the transporting of a phospholipid vesicle across the interface has not yet been reported, and the mechanism of this transporting is poorly understood.

To examine the transporting of a phospholipid vesicle, we conducted a microscopic observation of a water/oil/phospholipid macro phase-separation system; water and oil containing phospholipids are separated by a water/oil interface where the lipids are aligned with their hydrophilic parts facing the water phase. It is shown that a W/O CE formed in the oil phase is transported into the water phase and back into the oil phase across the interface using a micromanipulator.

## 2. Experimental Section

**2. 1. Materials.** MilliPore MilliQ was used as a solvent in the water phase. A phospholipid, 1,2-dioleoyl-sn-glycero-3-phosphatidylcholine (DOPC) and oil (mineral oil) were purchased from Wako Pure Chemicals and Nacalai Tesque, respectively. A fluorescent probe, Texas red 1,2-dihexadecanoyl-sn-glycero-3-phosphoethanolamine, triethylammonuim salt (Texas red DHPE), was purchased from Invitrogen. 1-10mM DOPC including 10μM Texas red DHPE was dissolved in oil or water by ultrasonication for 90 min at 50 ℃ and used within one week.

**2. 2. Transporting of a phospholipid vesicle across an water/oil interface.** We established a water/oil/phospholipid macro-phase separation system, as shown in Fig. 1; water and oil containing phospholipids are separated by a water/oil interface, where the lipids are aligned with their hydrophilic parts facing the water phase. We conducted observations with a phase-contrast and fluorescent microscope (Nikon TE-300) equipped with a micromanipulation system (Narishige) which enabled us to inject several picoliters of substance and/or to move the glass capillary with a precision of only a few micrometers.

3. Results

**3. 1. Preparation of W/O CE.** Figure 2 shows phase-contrast and fluorescent microscopic images, and a schematic image of water droplets injected from a glass capillary into the oil phase containing phospholipids. A cell-sized water droplet in the oil phase spontaneously formed a sphere due to surface tension. The lipids were distributed at the water/oil boundary immediately after injection, accompanied by the formation of a stable phospholipid vesicle as W/O CE. The size of the water droplet can be controlled by the injection pressure or the diameter of the hole of the glass capillary.

**3. 2. Transporting of the lipid vesicle between the water and oil phases.** Figure 3 shows phase-contrast microscopic images and a schematic image of the procedure by which a phospholipid vesicle that formed in the oil phase was moved between the water and oil phases across the water/oil interface by micromanipulation. The water droplet injected in the oil phase tended to attach to the glass capillary and was transferred into the water phase across the interface while maintaining a closed vesicle formation, as shown in Fig. 3 A. The lipid vesicle in the water phase was transported into the oil phase across the interface by micromanipulation, as shown in Fig. 3 B, and then transferred back to the water phase, as shown in Fig. 3 C and D. There were two patterns in the transporting from the oil phase into the water phase; either the water/oil interface followed the lipid vesicle as in Fig. 3 A and D, or it did not, as in Fig. 3 C. We confirmed that the repetitive transporting between water and oil phases can be easily attained by use of the micromanipulator. We then intentionally broke the lipid vesicle in the water phase by swinging the vesicle around with micromanipulator, as shown in Fig. 3 E. Oil included in the lipid vesicle in the water phase appeared after the collapse of the lipid vesicle and moved toward the oil phase.

**3. 3. Spontaneous transformation of the vesicle accompanied by transfer across the water/oil interface.** We found that in the oil phase, phospholipids spontaneously form closed vesicles, and that when the lipid vesicle contacts with the water/oil interface, it spontaneously moves into the water phase across the interface. Figure 4 shows phase-contrast microscopic images and a schematic image of phospholipid vesicles in the oil phase spontaneously moving into the water phase.

## 4. Discussion

When a water droplet was placed in the oil phase containing phospholipids, the lipids would align with their hydrophilic parts facing the water phase at the water/oil boundary and spontaneously form a monolayer phospholipid vesicle as W/O CE. The lipid vesicle that was transported into the water phase across the water/oil interface by micromanipulation would add an outer monolayer at the interface and transform into a bilayer phospholipid vesicle in the water phase. The lipid vesicle in the water phase contained a little oil between the acyl chains of phospholipids and the volume of the oil would depend on the kinetics of transporting. If the oil is uniformly distributed between the acyl chains, the thickness of the membrane was much larger than that of CL, which is about 5 nm. The oil that remained in the lipid vesicle had to be removed for the formation of a complete CL. In relation to the present result, it has recently been reported that double emulsions were formed across water/oil interface[28], where the volatile oil of the middle layer left after formation of a double emulsion accompanied by the formation of polymerosomes. The size of the water droplet in the oil phase can be controlled by the injection pressure or by the diameter of the hole of the glass capillary. Thus, it is quite possible to control the size of the CL formed by transporting a lipid vesicle in the oil phase into the water phase. This technique also enables us to include a substance in the interior of the CL.

The lipid vesicle assumes different structures in the water and oil phases, respectively. In the water phase, the hydrophilic parts of the outermost layer of the lipid vesicle face the water phase, while in the oil phase the acyl chains of that face the oil phase. When the lipid vesicle in the water phase is transported into the oil phase, the lipid vesicle gains one more outer layer at the water/oil interface and changes into a multilayer, probably trilayer, phospholipid vesicle, as shown in Fig. 3 B. The lipid vesicle in the oil phase could be placed back in the water phase, as shown in Fig. 3 C and D. In Fig. 3 C, the trilayer lipid vesicle would lose its outermost monolayer at the interface and be transported into the water phase while changing into a bilayer CL. On the other hand, in Fig. 3 D, the trilayer lipid vesicle lost its outermost bilayer while changing into W/O CE, and then gained an outer monolayer at the interface. We consider that this difference in the procedure for the transition determines whether the water/oil interface follows the lipid vesicle during transporting over the interface. Thus, these results suggest that the transition of W/O CE into CL across the interface requires a force such as drafting by micromanipulation. On the other hand, a trilayer lipid vesicle is expected to spontaneously lose its outer monolayer while changing into CL. The lipid vesicle transported from the water phase into the oil phase had a multilayer configuration every time. We could break the multilayer configuration at any time and form a monolayer as W/O CE by contacting the interface and remaining in place for several seconds using the micromanipulator, as shown in Fig. 3. D.

Meanwhile, the lipid vesicle that spontaneously formed in the oil phase was probably an inverted CL, a bilayer phospholipid vesicle with the hydrophilic parts of the phospholipids face-to-face. An inverted CL spontaneously lost its outer monolayer at the water/oil interface and changed into a monolayer phospholipid vesicle as an oil-in-water cell-sized emulsion (O/W CE). This result corresponds to the transition of the trilayer lipid vesicle in the oil phase into CL in Fig. 3 C, where the lipid vesicle loses the outer monolayer at the interface. We expect that the outer layer was absorbed into the interface accompanied by formation of micelles in the water phase, where before and after the transition the decrease in the chemical potential of phospholipids of the outer layer would be grater than the increase in free energy due to the surface energy of O/W CE. Figure 5 shows phase-contrast images and a schematic image of the transition of the lipid vesicle in a chamber like Fig. 1 (b) and a plot of the time development of $\theta$ during this transition, where the effect of the cover glass on the transition is less than that in Fig. 4 using a chamber as in Fig. 1 (a) with the water/oil interface in contact with the cover glass. The inverted CL lost its outer layer with a nearly constant velocity, $d\theta/dt$. It has been discussed in a previous report that a multi-bilayer phospholipid vesicle lost its outer bilayer with an almost constant velocity due to a surface tension induced by the attached solid-liquid interface.[29] We consider that this calculation can be directly applied to our result. We consider the layer density of the vesicle $\ell(\theta,t)$ as an order parameter, the time development of which would be given as

$$\frac{\partial \ell(\theta)}{\partial t} = -L \frac{\delta F(\ell(\theta))}{\delta \ell} \tag{1}$$

$$F(\ell(\theta)) = \int d\theta \left( f(\ell(\theta)) + \frac{C}{2} (\nabla \ell(\theta))^2 \right) \tag{2}$$

where $F(\ell)$ is the Ginzburg-Landau-type free energy function. In equation (2), the first term in the integral is the free energy density on $\theta$, and the second term describes the effect of a spatial gradient on the order parameter, which corresponds to the line energy due to the highly curved edge of the outer layer of the lipid vesicle during the transition. We consider $f(\ell)$ as quartic to describe a bistable model; metastable states $\ell_1$ and $\ell_2$, respectively, before and after the transition exist via unstable state $\ell_3$ during the transition, and can be written as

$$\frac{\delta f(\ell)}{\delta \ell} = a(\ell - \ell_1)(\ell - \ell_2)(\ell - \ell_3) \tag{3}$$

When a wave front of the transition exists at $\theta'$, $\ell(\theta, t)$ can be written as $\ell(\theta - v(\theta')t)$ where $v(\theta)$ is the velocity of the transition. Under the boundary condition $\ell(-\infty) = \ell(\theta_1)$ and $\ell(\infty) = \ell(\theta_2)$, from (1)-(3) $v(\theta)$ is calculated as

$$v(\theta) \cong \frac{L}{r}\sqrt{\frac{aC}{2}}(\ell_1 + \ell_2 - 2\ell_0) - \frac{CL}{r^2}\cot\theta \tag{4}$$

indicating that the transition occurs with almost constant velocity, except at $\theta \sim 0$ and $\pi$.

## 5. Conclusion

We conducted microscopic observations of a water/oil/phospholipid macro phase-separation system; water and oil containing phospholipids are separated by a water/oil interface where the lipids are aligned with their hydrophilic parts facing the water phase. We found that a cell-sized phospholipid vesicle can be transported between the water and oil phases with the use of a micromanipulator. The lipid vesicle takes an appropriate form in the water and oil phases, respectively, by losing or adding the outermost layer at the interface. We also found that a phospholipid vesicle, probably a bilayer vesicle, in the oil phase spontaneously loses its outer layer and transforms to a monolayer vesicle. These phospholipid vesicles have different characteristics according to the number of layers and the conditions around the vesicle, i.e. water or oil phase. This technique makes phospholipid vesicles a more useful tool as a cell model and microreactor, and provides a prospective method for preparing CLs which enables us to encapsulate biochemical substances within the interior and/or to control the size of CLs.

**6. Acknowledgement**

This work was supported by a Grant-in-Aid for the 21st Century COE "Center for Diversity and Universality in Physics" and a Grant-in-Aid for Scientific Research on Priority Areas (No.17076007) "System Cell Engineering by Multi-scale Manipulation" from the Ministry of Education, Culture, Sports, Science and Technology of Japan. M. Hase was financially supported by Kyoto University Venture Business Laboratory. A. Yamada was financially supported by a Sasagawa Scientific Research Grant from The

Japan Science Society. T. Hamada was supported by a Research Fellowship from the Japan Society for the Promotion of Science for Young Scientists (No. 16000653). We thank Prof. N. Kumazawa, Faculty of Engineering, Ibaraki University, for his kind advice on the experimental procedures.

# 7. References

(1) Bretscher, M. S. The molecules of the cell membrane. Sci. Am. **1985**, 253, 100–109

(2) Walde, P.; Ichikawa, S. Enzymes inside lipid vesicles: preparation, reactivity and applications. Biomol. Eng. **2001**, 18, 143-177

(3) Monnard, P. A. Liposome-entrapped polymerases as models for microscale/nanoscale bioreactors. J. Membr. Biol. **2003**, 191, 87-97

(4) Ourisson, G..; Nakatani, Y. Chem. Biol. **1994**, 1, 11-23

(5) Fendler, J. H. "Membrane Mimetic Chemistry," John Wiley & Sons, New York (1982)

(6) Luisi, P. L.; Wakde, P. Giant Vesicles, John Wiley & Sons Ltd, 2000

(7) Hotani, H.; Nomura, F.; Suzuki, Y. Giant liposomes: from membrane dynamics to cell morphogenesis. Curr. Opin. Colloid Interface Sci. **1999**, 4, 358-368

(8) Limozin, L.; Roth, A.; Sackmann, E. Microviscoelastic moduli of biomimetic cell envelopes. Phys. Rev. Lett. **2005**, 95, 178101

(9) Sato, Y.; Nomura, S. M.; Yoshikawa, K. Enhanced uptake of giant DNA in cell-sized liposomes. Chem. Phys. Lett. **2003**, 380, 279-285

(10) Michel, M.; Winterhalter, M.; Darbois, L.; Hemmerle, J.; Voegel, J. C.; Schaaf, P.; Ball, V. Giant liposome microreactors for controlled production of calcium phosphate

crystals. Langmuir **2004**, 20, 6127-6133

(11) Noireaux, V.; Libchaber, A. A vesicle bioreactor as a step toward an artificial cell assembly. Proc. Natl. Acad. Sci. USA **2004**, 101, 17669-17674

(12) Nomura, S. M.; Tsumoto, K.; Hamada, T.; Akiyoshi, K.; Nakatani, Y.; Yoshikawa, K. Gene expression within cell-sized lipid vesicles. ChemBioChem **2003**, 4, 1172-1175

(13) Chui, D. T.; Wilson, C. F.; Ryttsén, F.; Strömberg, A.; Farre, C.; Karlsson, A.; Nordholm, S.; Gaggar, A.; Moki, B. P.; Moscho, A.; Garza-lópez, R. A.; Orwar, O.; Zare, R. N. Chemical transformations in individual ultrasmall biomimetic containers. Science **1999**, 283, 1892-1895

(14) Kulin, S.; Kishore, R.; Helmerson, K.; Locascio, L. Optical manipulation and fusion of liposomes as microreactors. Langmuir **2003**, 19, 8206-8210

(15) Szoka, F.; Papahadjopoulos, D. Procedure for preparation of liposomes with large internal aqueous space and high capture by reverse-phase evaporation. Proc. Natl. Acad. Sci. U.S.A. **1978**, 75, 4194

(16) Deamer, D.; Bangham, A. D. Large volume liposomes by an ether vaporization method. Biochim. Biophys. Acta **1976**, 443, 629

(17) Matsumoto, S.; Kohda, M.; Murata, S. Preparation of lipid vesicles on the basis of a technique for providing W/O/W emulsions. J. Colloid Interface Sci. **1977**, 62, 149

(18) Papahadjopoulos, D.; Vail, W. J.; Jacobson, K.; Poste, G. Cochleate lipid cylinders-formation by fusion of unilamellar lipid vesicles. Biochim. Biophys. Acta **1975**, 329, 483-491

(19) Hub, H. H.; Zimmermann, U.; Ringsdorf, H. Preparation of large unilamellar vesicles. FEBS Lett. **1982**, 140, 254

(20) Mueller, P.; Chien, T. F.; Rudy, B. Formation and properties of cell-size lipid bilayer vesicles. Biophys. J. **1983**, 44, 375

(21) Magome, N.; Takemura, T.; Yoshikawa, K. Spontaneous formation of giant liposomes from neutral phospholipids. Chem. Lett. **1997**, 205-206

(22) Hauser, H. Some aspects of the phase behaviour of charged lipids. Biochim. Biophys. Acta **1984**, 772, 37

(23) Kabalnov, A.; Wennerström, H. Macroemulsion stability: the oriented wedge theory revisited. Langmuir **1996**, 12, 276-292

(24) Pietrini, A. V.; Luisi, P. L. Cell-free protein synthesis through solubilisate exhange in water/oil emulsion compartments. ChemBioChem **2004**, 5, 1055-1062

(25) Griffiths, A. D.; Tawfik, D. S.; Directed evolution of an extremely fast phosphotriesterase by in vitro compartmentalization. EMBO J. **2003**, 22, 24-35

(26) Katsura, S.; Yamaguchi, A.; Inami, H.; Matsuura, S.; Hirano, K.; Mizuno, A.

Indirect micromanipulation of single molecules in water-in-oil emulsion. Electrophoresis **2001**, 22, 289-293

(27) Sophie, P.; Barbara, J.; Frisken,; Weitz, D. A. Production of unilamellar vesicles using an inverted emulsion. Langmuir **2003**, 19, 2870-2879

(28) Lorenceau, E.; Utada, A. S.; Link, D. R.; Cristobl, G.; Joanicot, M.; Weitz, D. A. Generation of polymerosomes from double-emulsions. Langmuir **2005**, 21, 9183-9186

(29) Hamada, T.; Yoshikawa, K. Peeling kinetics of giant multilamellar vesicles on a solid-liquid interface. Chem. Phys. Lett. **2004**, 396, 303-307

**Figure Captions**

Fig 1: Schematic image of a water/oil/phospholipid macro phase-separation system. Water and oil containing phospholipids are separated either horizontally (a), or vertically (b) in a chamber made from silicone rubber by an water/oil interface where the lipids are aligned with their hydrophilic parts facing the water phase.

Fig. 2: Preparation of W/O CE in the oil phase containing phospholipids by micromanipulation. (a) Phase-contrast microscopic images of a procedure in which a cell-sized water droplet was injected into the oil phase from a glass capillary. The size of the water droplet was controlled by the injection pressure or the diameter of the hole of the glass capillary. Several sizes of W/O CEs were formed in the oil phase. (b) Fluorescent microscopic image of W/O CE immediately after injection of the water droplet. (c) Schematic drawing of the procedure used to prepare a W/O CE

Fig. 3: Phase-contrast microscopic images and schematic image of the procedure in which a phospholipid vesicle was transported between the water and oil phases by micromanipulation. A-D: W/O CE formed in the oil phase was attached to the glass capillary and transferred between the water and oil phases, where a phospholipid vesicle

was formed, which assumed different structures in the water and oil phases, respectively, by losing and adding on outer layer at the water/oil interface. E: The oil that was contained in the lipid vesicle in the water phase appeared after collapse of the lipid vesicle and was headed toward the oil phase. F: Schematic representation of the transporting of a phospholipid vesicle between the water and oil phases.

Fig. 4: Phase-contrast microscopic images and schematic image of the procedure by which a phospholipid vesicle in the oil phase spontaneously moved into the water phase across the water/oil interface. (a) Phospholipid vesicle in the oil phase. (b) The lipid vesicle formed in the oil phase spontaneously moved to the water phase across the water/oil interface. (c) Schematic representation of the transition of the lipid vesicle in the oil phase into the water phase across the interface

Fig. 5: Phospholipid vesicle in the oil phase spontaneously moved into the water phase across the water/oil interface, where the water and oil phases were separated vertically. The focal point of the microscope was set at the water phase and the lipid vesicle in the oil phase was observed in the same frame. (a),(b) Phase-contrast microscopic images and schematic image, respectively, of a bilayer phospholipid vesicle in the oil phase

during transformation into a monolayer. (c) Plot of the time development of θ during the transformation. The lipid vesicle in the oil phase lost its the outer layer with a nearly constant velocity, dθ/dt.

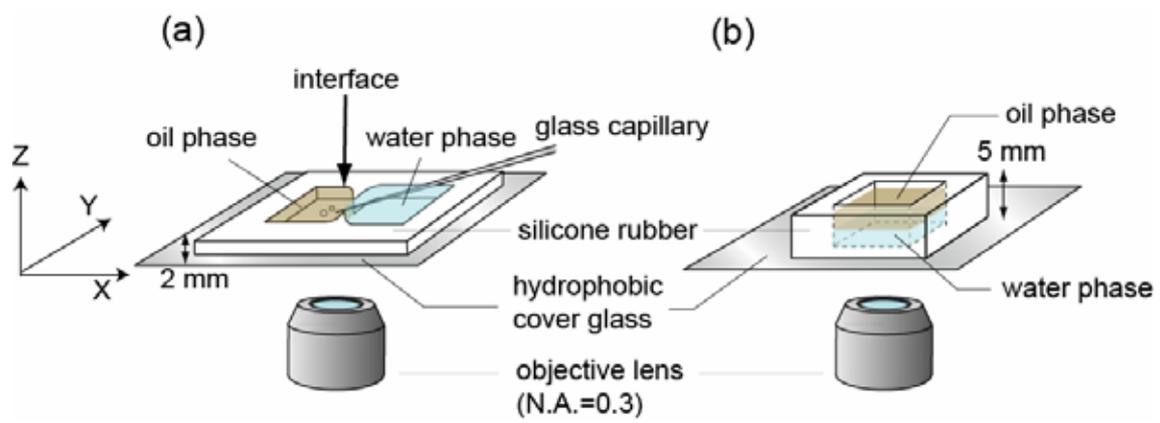

Figure 1

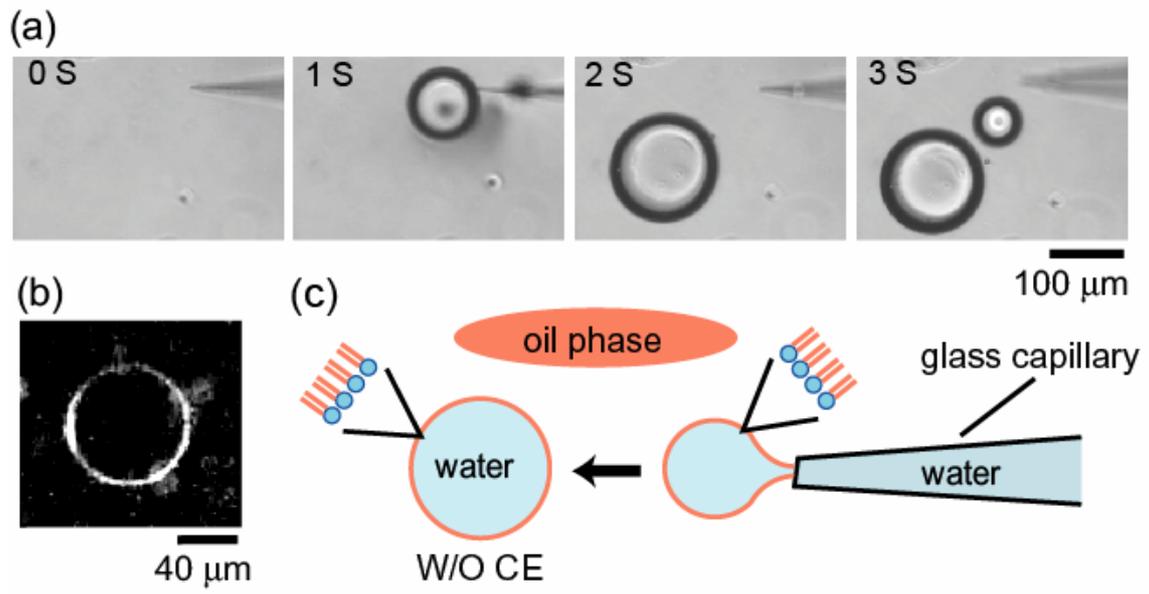

Figure 2

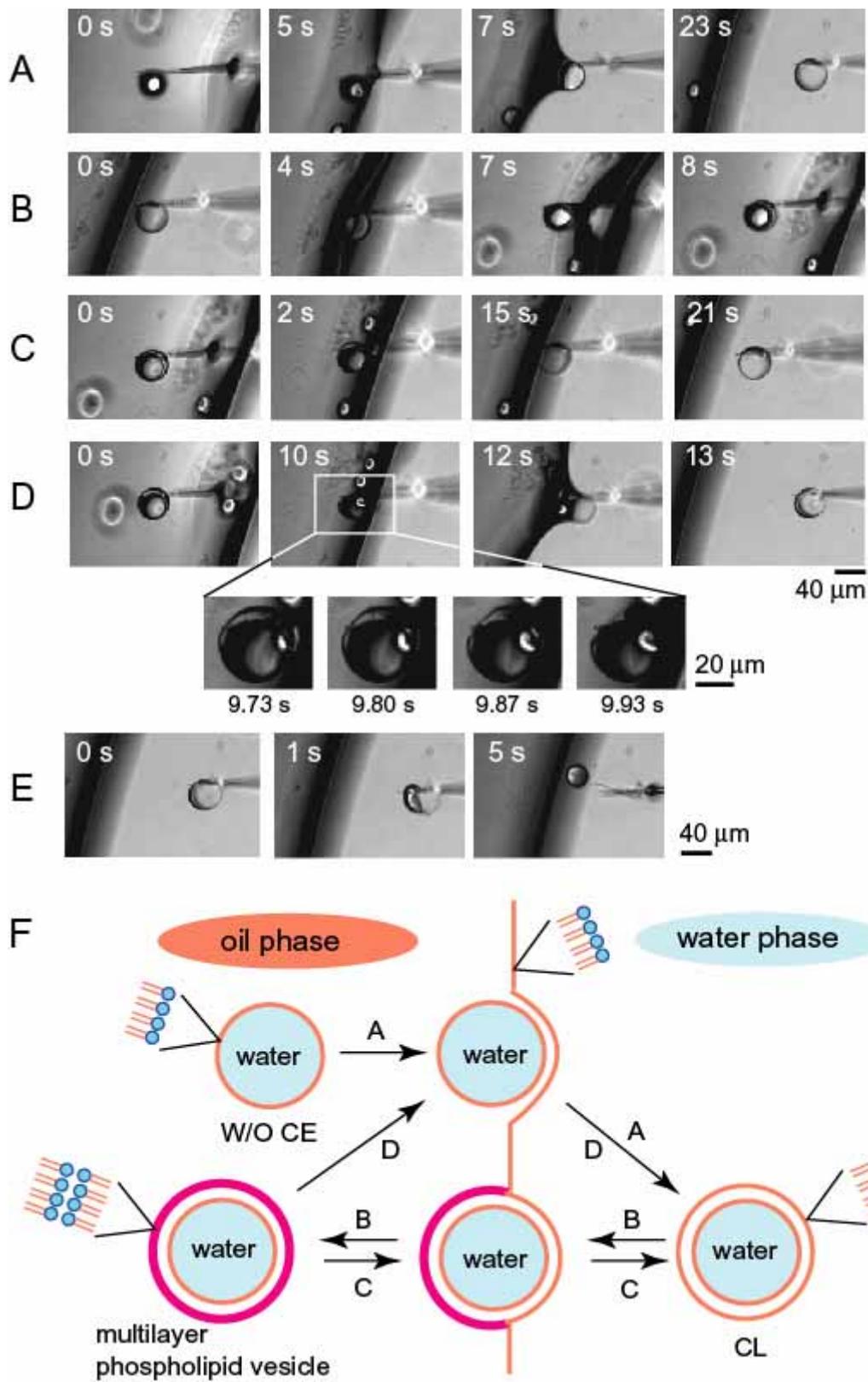

Figure 3

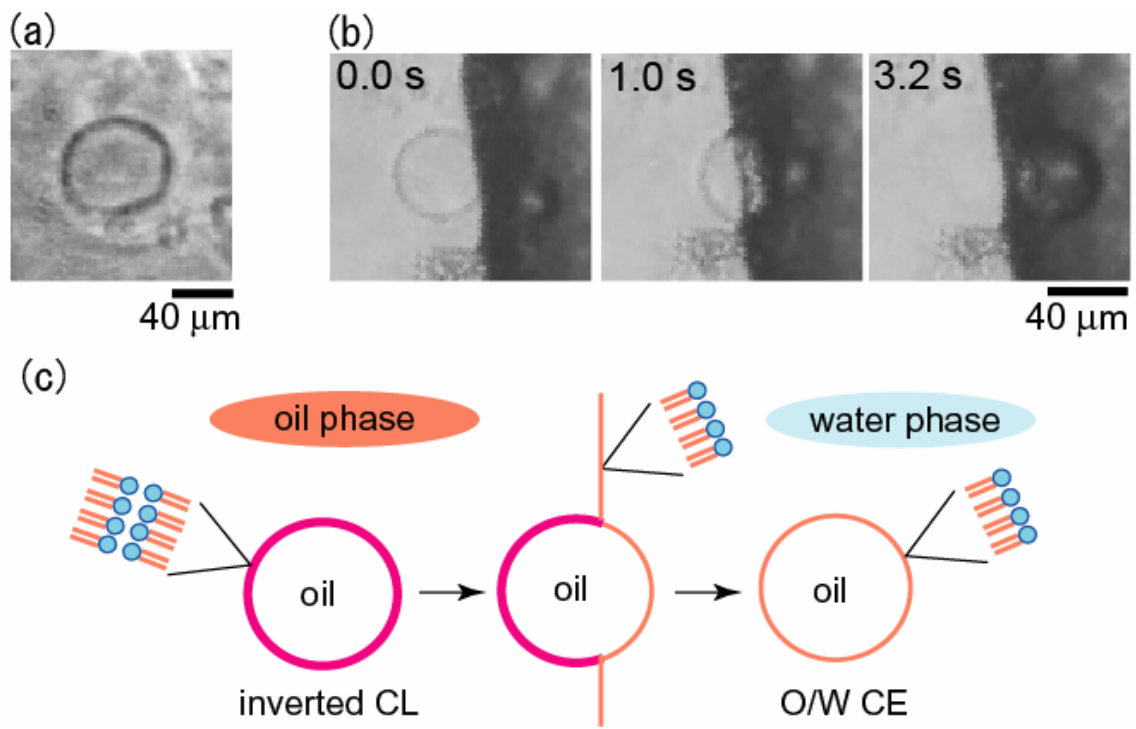

Figure 4

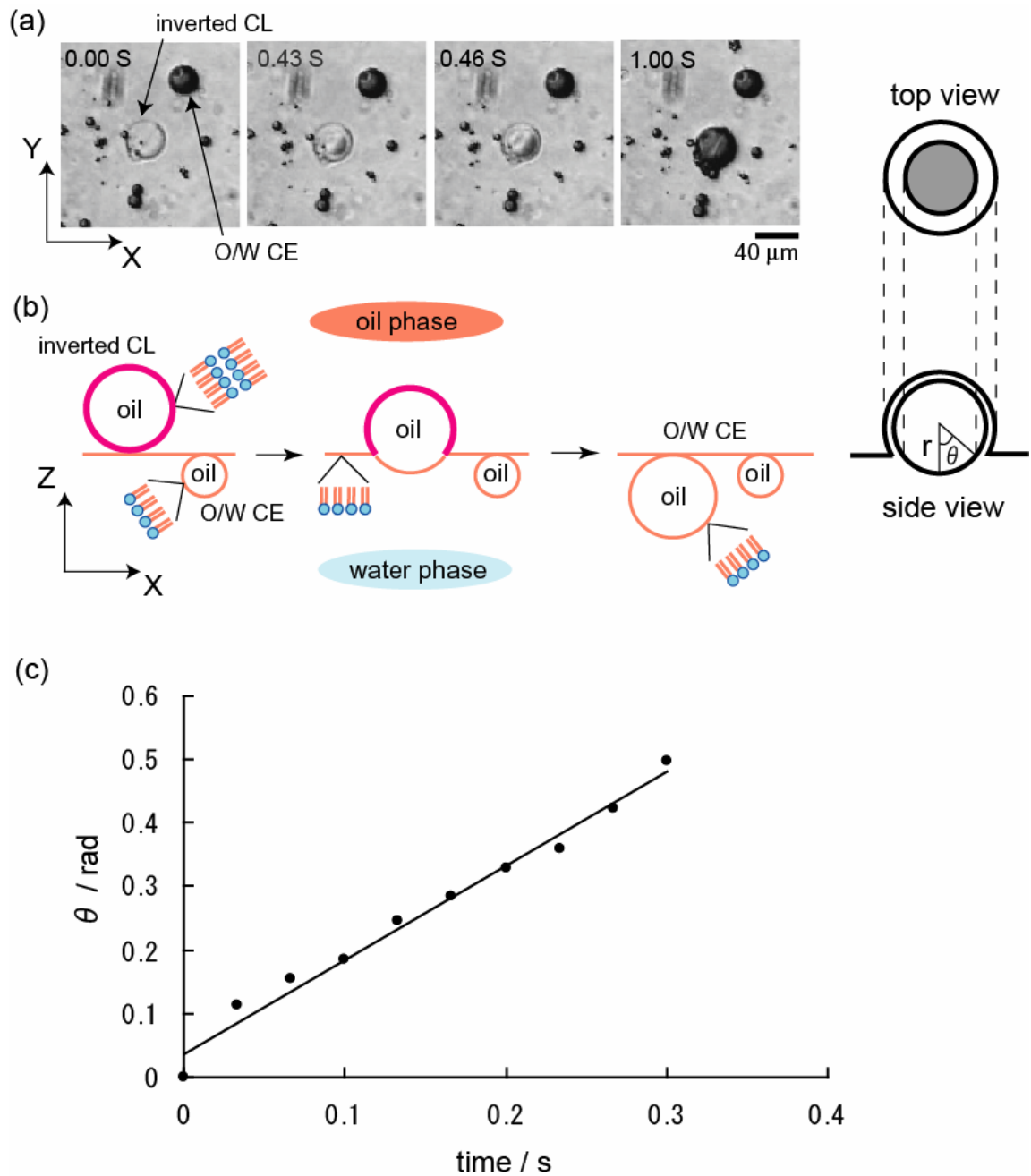

Figure 5

**Table of Contents Graphic**

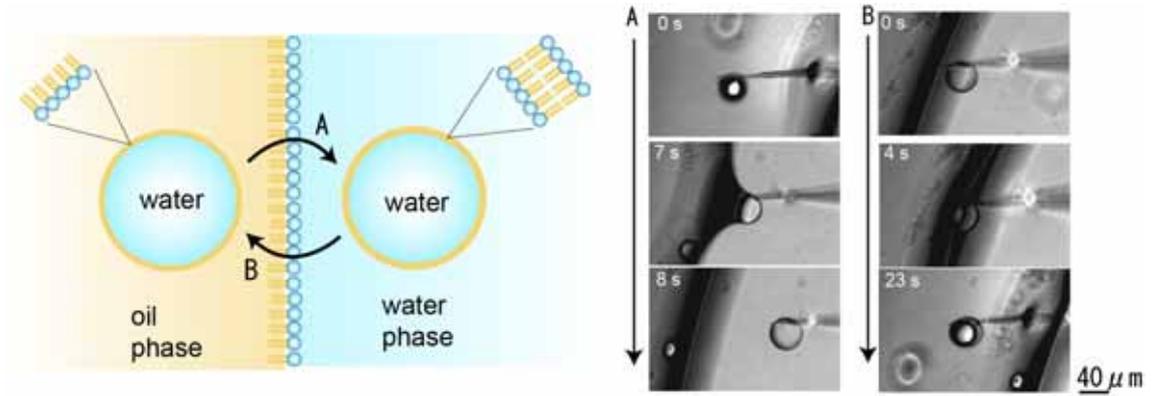